\documentstyle[11pt,iau177,twoside,psfig]{article}

\begin{document}


\title{Small-eccentricity binary pulsars and relativistic gravity}

\author{Norbert Wex}

\affil{Max-Planck-Institut f\"ur Radioastronomie, 
       Auf dem H\"ugel 69, 53121 Bonn, Germany}

\begin{abstract}
Small-eccentricity binary pulsars with white dwarf companions provide
excellent test laboratories for various effects predicted by alternative
theories of gravity, in particular tests for the emission of gravitational
dipole radiation and the existence of gravitational Stark effects. We will
present new limits to these effects. The statistical analysis presented here,
for the first time, takes appropriately care of selection effects.
\end{abstract}

\vspace{-3ex}

\section{Introduction}

The majority of binary pulsars is found to be in orbit with a white-dwarf
companion. Due to the mass transfer in the past, these systems have very small
orbital eccentricities and, therefore, neither the relativistic advance of
periastron nor the Einstein delay were measured for any of these binary
pulsars. In fact, the only post-Keplerian parameters measured with reasonable
accuracy for a small-eccentricity binary pulsar are the two Shapiro parameters
in case of PSR B1855+09 (Kaspi {\it et al.\ }1994). However, since the orbital
period of this system is 12.3 days, the expected gravitational wave damping of
the orbital motion is by far too small to be of any importance for timing
observations and, consequently, there is no third post-Keplerian parameter
which would allow the kind of test conducted in double-neutron-star systems
(Damour \& Taylor 1992). On the other hand, many alternative theories of
gravity, tensor-scalar theories for instance, predict effects that depend
strongly on the difference between the gravitational self energy per unit mass
($\epsilon\equiv E^{{\rm grav}}/mc^2$) of the two masses of a binary systems
(Will 1993; Damour \& Esposito-Far\'ese 1996ab). While this difference in
binding energies is comparably small for double-neutron-star systems, it is
large in neutron star-white dwarf systems since for a white dwarf
$\epsilon\sim 10^{-4}$ while for a 1.4$M_\odot$ neutron star $\epsilon \approx
0.15$.

\vspace{-1ex}

\section{Gravitational dipole radiation}

Unlike general relativity, many alternative theories of gravity predict the
presence of all radiative multipoles --- monopole and dipole, as well as
quadrupole and higher multipoles (Will 1993). For binary systems scalar-tensor
theories, for instance, predict a loss of orbital energy which at highest
order is dominated by scalar dipole radiation. As a result, the orbital
period, $P_b$, of a circular binary system should change according to
\begin{equation} 
   \dot P_b^{({\rm dipole})} \simeq -\frac{4\pi^2G_*}{c^3P_b} \;
                            \frac{m_pm_c}{m_p+m_c} \;
                            (\alpha_p-\alpha_c)^2 \;,
\end{equation}
where $m_p$ and $m_c$ denote the mass of the pulsar and its companion,
respectively, $G_*$ is the `bare' gravitational constant and $c$ the speed of
light. The total scalar charge of each star is proportional to its mass and
its `effective coupling strength' $\alpha(\epsilon)$ (Damour \&
Esposito-Far\'ese 1996b).  For a white dwarf companion $|\alpha_c|\ll 1$ and
thus the expression $(\alpha_p-\alpha_c)^2$ in equation (1) can be of the
order one if the pulsar develops a significant amount of scalar charge. In
this case the gravitational wave damping of the orbit is completely dominated
by the emission of gravitational dipole radiation.

PSR J1012+5307 is a 5.3 ms pulsar in a 14.5 h circular orbit with a low mass
white-dwarf companion. Since its discovery in 1993 (Nicastro {\it et al.\
}1995) this pulsar has been timed on a regular basis using the Jodrell Bank
76-m and the Effelsberg 100-m radiotelescope, sometimes achieving a timing
accuracy of 500 ns after just 10 min of integration (Lange {\it et al.\ },
this conference). In addition, the white-dwarf companion appears to be
relatively bright ($V=19.6$) and shows strong Balmer absorption lines. Based
on white dwarf model calculations, a companion mass of $m_c=0.16\pm0.02$ and a
distance of $840\pm90$ pc was derived (van Kerkwijk {\it et al.\ }1996,
Callanan {\it et al.\ }1998).  Further, a reliable radial velocity curve for
the white dwarf has been extracted, which then, in combination with the pulsar
timing information, gave a mass for the pulsar of $m_p = 1.64 \pm 0.22$. Since
~$\dot P_b = (0.1\pm1.8)\times 10^{-13}$ for this binary system, we find from
equation (1)
\begin{equation}
   |\alpha_p|<0.02 \quad \mbox{(95\% C.L.)}
\end{equation}
Simulations show, that this value should improve by a factor of five within
the next three years (Lange {\it et al.}, in prep.).

\vspace{-1ex}

\section{Gravitational Stark effects}

\subsection{Violation of the strong equivalence principle}

The {\em strong equivalence principle (SEP)} requires the universality of free
fall of all objects in an external gravitational field regardless of their
mass, composition and fraction of gravitational self-energy. While all metric
theories of gravity share the property of universality of free fall of test
particles (weak equivalence principle), many of them, which are considered as
realistic alternatives to general relativity, predict a violation of the
SEP. A violation of the SEP can be understood as an inequality between the
gravitational mass, $m_g$, and the inertial mass, $m_i$, which can be written
as function of $\epsilon$:
\begin{equation}
   m_g/m_i \equiv 1 + \delta(\epsilon) 
      = 1 + \eta \epsilon + {\cal O}(\epsilon^2) \;.
\end{equation}
While the analysis of lunar-laser-ranging data tightly constrains the
`Nordtvedt parameter' $\eta$ (M\"uller {\it et al.\ }1997) it indicates
nothing about a violation of the SEP in strong-field regimes, i.e.\ terms of
higher order in $\epsilon$, due to the smallness of $\epsilon$ for
solar-system bodies. For neutron stars, however, $\epsilon \sim 0.15$ and thus
binary-pulsars with white-dwarf companions ($\epsilon \sim 10^{-4}$) provide
ideal laboratories for testing a violation of the SEP due to nonlinear
properties of the gravitational interaction (Damour \& Sch\"afer 1991).

In case of a violation of the SEP the eccentricity vector of a
small-eccentricity binary-pulsar system exposed to the external gravitational
field of the Galaxy, ${\bf g}$, is a superposition of a constant vector ${\bf
e}_F$ and a vector ${\bf e}_R$ which is turning in the orbital plane with the
rate of the relativistic advance of periastron. The `induced' eccentricity
${\bf e}_F$ points into the direction of the projection of the Galactic
acceleration onto the orbital plane, ${\bf g}_\perp$, and $e_F \propto
(\delta_p-\delta_c) P_b^2 g_\perp$.  However, neither the length of ${\bf
e}_R$ nor its rotational phase $\theta$ are known quantities. We therefore
have to proceed as follows.  Given a certain
$(\delta_p-\delta_c)\simeq\delta_p$, i.e.~a certain $e_F$ for a given binary
pulsar, the observed eccentricity $e$ sets an upper limit to $|\theta|$ which
is independent of $e_R$: $\sin|\theta|<e/e_F$ for $e<e_F$ and $|\theta|\le\pi$
for $e \ge e_F$ (Wex 1997). We now have to calculate an upper limit for
$\theta$ for every observed small-eccentricity binary pulsar and compare the
result with Monte-Carlo simulations of a large number of (cumulative)
distributions for the (uniformly distributed) angle $\theta$.  This way, by
counting the number of simulated distributions which are in agreement with the
`observed' limits, one obtains the confidence level with which a certain
$\delta_p$ is excluded.  As a safe upper limit for $|\delta_p|$ we find
\begin{equation}
   |\delta_p| < 0.009 \quad \mbox{(95\% C.L.)}
\end{equation}
Note, in order to calculate $e_F$ for a given binary system, we need also the
masses of pulsar and companion and the location and orientation of the binary
system in the Galaxy. If there are no restrictions from timing and optical
observations, the pulsar masses were assumed to be uniformly distributed in
the range $1.2M_\odot<m_p<2M_\odot$, the companion masses were taken from
evolutionary scenarios (Tauris \& Savonije 1999), and the pulsar distances
were estimated using the Taylor-Cordes model assuming a typical error of 25\%
(Taylor \& Cordes 1993). Finally, the orientation of the ascending node in the
sky, which is an unobservable parameter for all our binary system, was treated
as variable which is uniformly distributed between 0 and $2\pi$.

\subsection{Violation of local Lorentz invariance and conservation laws}

If gravity is mediated in part by a long-range vector field or by a second
tensor field one expects the global matter distribution in the Universe to
select a preferred frame for the gravitational interaction (Will \& Nordtvedt
1972). At the post-Newtonian level, gravitational effects associated with such
a {\em violation of the local Lorentz invariance} of gravity are characterized
by two theory dependent parameters $\alpha_1$ and $\alpha_2$.  
%
%
If $\alpha_1$ were different from zero, the eccentricity of a binary system
which moves with respect to the global matter distribution in the Universe
would suffer a secular change similar to a violation of the SEP.  This time,
$|e_F| \propto \alpha_1 |m_p-m_c| P_b^{1/3} w_\perp$ where ${\bf w}$ denotes
the velocity of the binary system with respect to the preferred frame,
i.e.~the cosmic microwave background.  Again, we can perform a Mote-Carlo
analysis as outlined in the previous section to derive the upper limit
\begin{equation}
   |\alpha_1| < 1.2\times 10^{-4} \quad \mbox{(95\% C.L.)}
\end{equation}
This limit is slightly better than the limit obtained from lunar-laser-ranging
data (M\"uller {\it et al.\ }1996) and, more importantly, also holds for
strong gravitational-field effects which could occur in the strong-field
regions of neutron stars. Due to its small eccentricity, $e<1.7\times 10^{-6}$
(95\% C.L.), and high velocity with respect to the cosmic microwave background
($w \approx 500$ km/s), PSR J1012+5307 turns out to be the most important
binary system for this kind of analysis (Lange {\it et al.\ }, in prep). While
for PSR J1012+5307 also the radial velocity of the system is known from
spectroscopic observations of the white dwarf companion, for all the other
binary pulsars no radial velocity information is available and we have to
assume an isotropic probability distribution for the 3-d velocity.

In theories of gravity which violate the local Lorentz invariance and the
momentum conservation law, a rotating self-gravitating body will suffer a
self-acceleration which is given by ${\bf a}_{{\rm self}} = -\frac{1}{3}
\alpha_3 \epsilon {\bf w} \times \mbox {\boldmath{$\Omega$}}$ (Nordtvedt \&
Will 1972), where $\alpha_3$ is a theory dependent parameter and $\mbox
{\boldmath{$\Omega$}}$ denotes the rotational velocity of the body. Again,
binary pulsars are ideal probes for this kind of self-acceleration effects
(Bell \& Damour 1996).  A careful analysis analogous to the previous analyses
(SEP, local Lorentz invariance) gives
\begin{equation}
   |\alpha_3| < 1.5 \times 10^{-19} \quad  \mbox{(95\% C.L.)}
\end{equation}

Note, the statistical tests for gravitational Stark effects presented here for
the first time appropriately take care of selection effects by simulating the
whole population, therefore, giving the first reliable limits for $\delta_p$,
$\alpha_1$, and $\alpha_3$.

\acknowledgements

I am grateful to Kenneth Nordtvedt for pointing out to me the problem of
selection effects related with binary-pulsar limits to gravitational Stark
effects. I thank Christoph Lange for numerous valuable discussions.

\end{document}